\pdfoutput=1
\documentclass[aps,prd,twocolumn,superscriptaddress,showpacs,preprintnumbers]{revtex4-1}
\usepackage[colorlinks=true, pdfstartview=FitV, linkcolor=red, citecolor=blue, urlcolor=black, pdftitle={},pdfauthor={Tomoya Hayata},pdfsubject={}, pdfkeywords={}]{hyperref}

\usepackage[dvipdfmx]{graphicx}
\usepackage{amssymb,bm}
\usepackage{amsthm,amsmath}
\usepackage{color}

\begin{document}
\preprint{RIKEN-QHP-87}
\title{Dynamical Mass Generation of Vector Mesons from QCD Trace Anomaly}
\author{Tomoya Hayata}
\affiliation{Department of Physics, The University of Tokyo, Tokyo 113-0031, Japan}
\affiliation{Theoretical Research Division, Nishina Center, RIKEN, Wako 351-0198, Japan}

\date{September 6, 2013}

\begin{abstract}

 Mass formulas for the vector mesons written in terms of the gluon condensate i.e.,
 the trace anomaly in quantum chromodynamics (QCD),
 are derived on the basis of finite energy QCD sum rules.
 We utilize sum rules with $s^n$ and $s^{n+1/2}$ weights,
 which relate the energy-weighted spectral sums
 to the vacuum expectation values of certain commutation relations.
 After evaluating the commutation relations,
 the sum rules with $s^n$ weights are reduced to
 the familiar ones obtained from the operator product expansion (OPE).
 On the other hand, the sum rules with $s^{n+1/2}$ weights cannot be derived from OPE. 
 They give new relations between the spectral sums
 and QCD vacuum fluctuations. 
 To derive simple mass formula, 
 we adopt the pole $+$ continuum Ansatz for the spectral function,
 and solve coupled equations given by the sum rules with $s^{0,1}$ weights and
 the new sum rule with $s^{1/2}$ weight.
 Application of our approach to the axial-vector meson is also discussed.

\end{abstract}

\pacs{11.55.Hx,12.38.Aw,12.38.Lg,14.40Be}

\maketitle

\section{Introduction}

 One of the major goals of quantum chromodynamics (QCD) is
 to understand the hadron spectra  
 from the non-perturbative dynamics of quarks and gluons.
 The method of QCD sum rules, which was originally proposed
 by Shifman, Vainshtein and Zakharov~\cite{Shifman:1978bx},
 is a useful theoretical tool for this problem and 
 has been widely used to relate the hadron dynamics
 to the QCD vacuum condensates~\cite{Shifman:2001,Narison:2002pw,Brambilla:2010cs}. 
 Also, the numerical simulation based on the lattice QCD has become a powerful tool
 to study the hadron spectroscopy from the first principles of QCD~\cite{Fodor:2012gf}.

 Since hadrons are the low energy collective modes in the QCD vacuum, 
 their dynamical scale is determined
 by the QCD scale parameter ($\Lambda_{\rm QCD}$) of the order of $200$ MeV, or equivalently,
 by the QCD vacuum condensates, which are scaled by $\Lambda_{\rm QCD}$.
 For example, the $\rho$-meson mass was estimated
 from the chiral condensate $\langle0|\bar{q}q|0\rangle\sim\Lambda^3_{\rm QCD}$
 on the basis of QCD sum rules approach~\cite{Shifman:1978bx,Krasnikov:1982ea}:
 $m_{\rho}\sim|\langle0|\bar{q}q|0\rangle|^{1/3}$.
 Here, so called ``vacuum saturation Ansatz'' was adopted 
 to evaluate the vacuum expectation values of the four-quark operators.
 On the other hand, the introduction of $\Lambda_{\rm QCD}$ through renormalization
 is closely related to
 the trace anomaly of QCD, namely,
 the gluon condensate
 $\langle0|\alpha_s G^{a\mu\nu}G^{a}_{\mu\nu}|0\rangle\sim\Lambda_{\rm QCD}^4$~\cite{'tHooft:2005ji}.
 Therefore, it would be possible that one derives the mass formulas of hadrons
 written in terms of the gluon condensate,
 and estimates the hadron masses from the gluon condensate without factorization Ansatz.

 In this paper, mass formulas for the vector mesons written in terms of the gluon condensate
 are derived on the basis of finite energy QCD sum rules.
 We start with the spectral decomposition of the current correlation function, and
 derive the sum rules for the vector current
 with $s^n$ and $s^{n+1/2}$ weights,
 in which the energy-weighted spectral sums
 are represented by the vacuum expectation values of certain commutation relations.
 After evaluating the commutation relations,
 the sum rules with $s^{n}$ weights are 
 reduced to the familiar ones obtained from the operator product expansion (OPE). 
 A detail of derivation will be discussed in the forthcoming publication~\cite{Hayata}.
 On the other hand, the sum rules with $s^{n+1/2}$ weights cannot be derived from OPE.  
 They give new relations, in which the spectral sums
 are represented by QCD vacuum fluctuations. 
 By adopting the standard pole $+$ continuum Ansatz for the spectral function,
 we solve the sum rules with $s^{0,1/2,1}$ weights, 
 and obtain the mass formulas of the vector mesons. 
 Our new sum rule with $s^{1/2}$ weight is utilized
 instead of the sum rule with $s^2$ weight usually used in the standard QCD sum rules approach.
 We will mention its relation to our approach.
 Application of our approach to the axial-vector meson is also discussed.

 \section{Finite energy QCD sum rules from commutation relations}

 Let us start with a time-ordered current correlation function in the vacuum:
\begin{eqnarray}
  \Pi_{\mu\nu}(q)&=& i\int\mathrm{d}^4x\;\mathrm{e}^{iqx}\langle0|\mathrm{T}[j_{\mu}(x)j_{\nu}(0)]|0\rangle .
\end{eqnarray}
 The dimensionless spectral function $\rho(q^2)$ is defined as $\rho(q^2)=-{\rm Im}\Pi_{\mu}^{\mu}(q)/(3q^2)$,
 which is represented by the spectral decomposition as
\begin{equation}
  \rho(q^2)= -\frac{1}{3q^2}\sum_{p}\;(2\pi)^4\delta^{(4)}(q-p)\langle0|j^{\mu}(0)|p\rangle\langle p|j_{\mu}(0)|0\rangle .
  \label{sp1}
\end{equation}
 Here, $|p\rangle$ is an eigenstate of the QCD Hamiltonian with the energy $p^0$ and three-momentum $\bm{p}$.
 From the above spectral function, we define the spectral sums with $s^{n=0,1,\ldots}$ and $s^{n+1/2=1/2,3/2,\ldots}$ weights
 up to the high-energy cutoff $M$ as
\begin{eqnarray}
  \int_{0}^{M^2}\frac{\mathrm{d}s}{2\pi}\;s^n\rho(s) &=&
  \int_{0}^{M}\frac{\mathrm{d}q^0}{2\pi}\;2(q^0)^{2n+1}\rho((q^0)^2) ,
\label{sum_odd1} \\
  \int_{0}^{M^2}\frac{\mathrm{d}s}{2\pi}\;s^{n+1/2}\rho(s)  &=&  
  \int_{0}^{M}\frac{\mathrm{d}q^0}{2\pi}\;2(q^0)^{2n+2}\rho((q^0)^2) ,
\label{sum_even1} 
 \end{eqnarray}
 where we take the three-momentum $\bm{q}$ zero.
 Since $\rho(s=q^2)$ is Lorentz invariant, we can choose a frame with $\bm{q}=0$
 to evaluate the sums without loss of generality.

 By substituting the spectral decomposition in Eq. (\ref{sp1}) into the right hand sides of
 Eqs.~(\ref{sum_odd1}) and~(\ref{sum_even1}), we obtain
\begin{eqnarray}
  &&\int_{0}^{M^2}\frac{\mathrm{d}s}{2\pi}\;s^n\rho(s) 
  \nonumber \\
  &&=-\frac{1}{3}\int\mathrm{d}^3x\;\langle0|\Bigl[[j_M^{\mu}(0,\bm{x}),\mathrm{H}]_{2n-1},j_{M\mu}(0,{\bf 0})\Bigr]|0\rangle ,
   \label{EWSR2} \\
  &&\int_{0}^{M^2}\frac{\mathrm{d}s}{2\pi}\;s^{n+1/2}\rho(s)  
  \nonumber \\
 && =\lim_{V\to\infty}\frac{2(-1)^{n+1}}{3V}\langle0|\Bigl(\int{\rm d}^3x\;[j_M^{\mu}(0,x),{\rm H}]_{n}\Bigr)^2|0\rangle,
  \label{sum_even2} 
\end{eqnarray}
 where ${\rm H}$ is the Hamiltonian of QCD obtained from Becchi-Rouet-Stora-Tyutin 
 invariant Lagrangian~\cite{Kugo:1977zq},
 brackets denote commutation relations, and
 $j_M^{\mu}=P_Mj^{\mu}P_M$ is the regularized current operator
 with $P_M=\sum_{p^0<M}\;|p\rangle\langle p|$ being a projection operator. 
 A short hand notation $[O,{\rm H}]_{m+1}=[[O,{\rm H}]_m,{\rm H}]$ is introduced with $[O,{\rm H}]_0=O$.
 By using the orthogonality of the energy-momentum eigenstates, we can check that $P_M$ satisfies 
 $P_M^2=P_M$, $[P_{M},{\rm H}]=0$.
 We utilized these identities to obtain Eqs.~(\ref{EWSR2}) and~(\ref{sum_even2}).
 Since $P_{M}$ commutes with ${\rm H}$,
 a commutation relation with ${\rm H}$ can be evaluated by using the equation of motion of $j^{\mu}$.
 Similar approach based on the commutation relations 
 is adopted to derive the energy-weighted sum rules for quantum many-body systems~\cite{Suzuki1984}.

 In the limit $M\rightarrow\infty$, both sides of the sum rules
 (the spectral sums and the vacuum expectation values of the commutation relations)
 in Eqs.~(\ref{EWSR2}) and~(\ref{sum_even2}) are in general ultraviolet (UV) divergent.
 Then, we need to carry out suitable subtraction of UV divergences
 to obtain finite sum rules in the limit $M\rightarrow \infty$.
 This can be achieved by subtracting the continuum spectral function
 $\rho^{{\rm pert.}}(s)$ from the left-hand-side
 and the associated perturbative evaluation of the vacuum expectation value 
 from the right-hand-side of Eq.~(\ref{EWSR2}):
\begin{eqnarray}
   && \int_{0}^{\infty}\frac{\mathrm{d}s}{2\pi}\;s^n(\rho(s)-\rho^{{\rm pert.}}(s)) 
  \nonumber \\
   &&=-\frac{1}{3}\int\mathrm{d}^3x\;\langle0|\Bigl[[j^{\mu}(0,\bm{x}),\mathrm{H}]_{2n-1},j_{\mu}(0,{\bf 0})\Bigr]|0\rangle_{\mathrm{NP}} , 
\label{eq:QCD-EWSR}
\end{eqnarray}
 where NP stands for the non-perturbative vacuum expectation value after such subtraction. 
 Now, $M$ can be taken to infinity, but its remnant appears
 as subtraction ambiguity: This causes the failure of the naive use of canonical commutation relations
 to evaluate the commutation relations in Eq.~(\ref{eq:QCD-EWSR}).
 The prescription of the commutation relations from the Hamiltonian formulation of QCD
 will be discussed in the forthcoming publication~\cite{Hayata}.
 After evaluating the commutation relations,
 the spectral sums can be represented by the QCD vacuum condensates
 and indeed correspond to the finite energy QCD sum rules obtained from OPE~\cite{Hayata}.

 By subtracting out divergent contribution from asymptotically free quarks and gluons
 in both sides of Eq.~(\ref{sum_even2}), we obtain finite sum rules with fractional powers:
\begin{eqnarray}
  &&  \int_{0}^{\infty}\frac{\mathrm{d}s}{2\pi}\;s^{n+1/2}\Bigl(\rho(s) -\rho^{\rm pert.}(s) \Bigr) 
  \nonumber \\
  &&  =\lim_{V\to\infty}\frac{2(-1)^{n+1}}{3V}\langle0|\Bigl(\int{\rm d}^3x\;[j^{\mu}(0,x),{\rm H}]_{n}\Bigr)^2|0\rangle_{\rm NP} .
  \label{sum_even4_2} 
\end{eqnarray}
 The spectral sums are not represented by the QCD vacuum condensates
 but by the correlations of certain composite operators.
 Therefore, they cannot be obtained from OPE.
 These new sum rules with $s^{n+1/2}$ weights give the relation between
 the low energy hadron resonances and the non-perturbative QCD vacuum fluctuations.
 To see this more clearly, let us consider the first moment with $\sqrt{s}$ weight:
\begin{equation}
 \int_{0}^{\infty}\frac{\mathrm{d}s}{2\pi}\;\sqrt{s}\Bigl(\rho(s) -\rho^{\rm pert.}(s) \Bigr)
 =\frac{2}{3}\chi_{\rm NP} .
 \label{sum_even5}
\end{equation}
 Here, the spectral sum is represented by a {\it generalized} susceptibility
 $\chi_{\rm NP}=-\langle0|Q^{\mu}Q_{\mu}|0\rangle_{\rm NP}/V\;(V\to\infty)$,
 with $Q^{\mu}=\int{\rm d}^3x\;j^{\mu}(0,\bm{x})$.
 We used this sum rule to derive the mass formulas for the vector mesons written in terms of the gluon condensate.

 \section{Mass formulas by gluon condensate}

 Let us consider the vector current with quantum numbers of the $\rho$ meson:
 $j^{\mu}_{\rho}=(\bar{u}\gamma^{\mu}u-d\gamma^{\mu}d)/2$.
 We take the following Ansatz for the spectral function:
\begin{equation}
 \rho_V(s)=F_{\rho}\delta(s-m_{\rho}^2)+\rho_V^{\rm pert.}(s)\Theta(s-s_{\rho0}),
 \label{pole}
\end{equation} 
 with $m_{\rho}$, $F_{\rho}$, $s_{\rho0}$ and $\rho_V^{\rm pert.}(s)$ being the mass, the pole residue,
 the continuum threshold, and the perturbative part of the spectral function, respectively.
 To represent these parameters by the gluon condensate, we solve
 the lowest three moments with $s^{n=0,1/2,1}$ weights, 
 which are given by, 
 in the linear order of $\alpha_s$ and the quark masses $m_{u,d}$,
\begin{eqnarray}
  &&\int_{0}^{\infty}\frac{\mathrm{d}s}{2\pi}\;\Bigl(\rho_V(s) -\rho_V^{\rm pert.}(s) \Bigr) = 0 ,
  \label{m1} \\
  &&\int_{0}^{\infty}\frac{\mathrm{d}s}{2\pi}\;\sqrt{s}\Bigl(\rho_V(s) -\rho^{\rm pert.}_V(s) \Bigr)
 =\frac{2}{3}\chi^{\rho}_{\rm NP} ,
 \label{m3} \\
 &&\int_{0}^{\infty}\frac{\mathrm{d}s}{2\pi}\;s\Bigl(\rho_V(s) -\rho_V^{\rm pert.}(s) \Bigr)
 \nonumber \\
 &&= \langle0|-\frac{m_u}{2}\bar{u}u-\frac{m_d}{2}\bar{d}d-\frac{\alpha_s}{24\pi}G^{a\mu\nu}G^a_{\mu\nu}|0\rangle_{\mathrm{NP}} ,
  \label{m2} 
\end{eqnarray}
 Here, we neglect the small $\alpha_s$-corrections in $\rho_V^{\rm pert.}(s)$,
 and take $\rho^{\rm pert.}_V(s)=1/(4\pi)$.
 We note that the chiral condensate $m_u\bar{u}u$ ($m_d\bar{d}d$) in Eq.~(\ref{m2})
 can be neglected numerically relative to the gluon condensate.
 First, by retaining only the gluon condensate
 and neglecting the generalized susceptibility ($\chi^{\rho}_{\rm NP}\rightarrow0$), we have
 $s_{\rho0}=9m_{\rho}^2/4$, $ F_{\rho}=9m_{\rho}^2/(16\pi)$, and 
\begin{eqnarray}
  m_{\rho} &=&\Bigl(\frac{32\pi}{27}\langle0|\alpha_sG^{a\mu\nu}G^a_{\mu\nu}|0\rangle_{\mathrm{NP}}\Bigr)^{\frac{1}{4}}
 \nonumber \\ 
  &\sim& 1.4\Bigl(\langle0|\alpha_sG^{a\mu\nu}G^a_{\mu\nu}|0\rangle_{\mathrm{NP}}\Bigr)^{\frac{1}{4}} . 
 \label{s3}
\end{eqnarray}
 Thsese are the main results of this paper.
 The operator $\alpha_sG^{a\mu\nu}G^a_{\mu\nu}$ is independent of
 a renormalization group point $\mu$ in the linear order of $\alpha_s$,
 so that we do not need to consider the explicit $\mu$-dependence of the gluon condensate.
 If we take $\langle0|\alpha_sG^{a\mu\nu}G^a_{\mu\nu}|0\rangle_{\mathrm{NP}}=7.0\times10^{-2}\;{\rm GeV}^4$,
 which was estimated in Ref.~\cite{Narison:2011xe},
 we have $m_{\rho}=710$ MeV.
 This is smaller only by $10$\% than the experimental value, $m_{\rho}=775$ MeV~\cite{Beringer:1900zz}.
 Furthermore, the coupling constant $g_{\rho}$ can be estimated from $F_{\rho}$ as
 $g_{\rho}^2/(4\pi)=8\pi/9\sim 2.8$, which is in agreement with the experimental value, 
 $g_{\rho}^2/(4\pi)=2.63$.

\begin{figure}
\centering
\includegraphics[scale=.65]{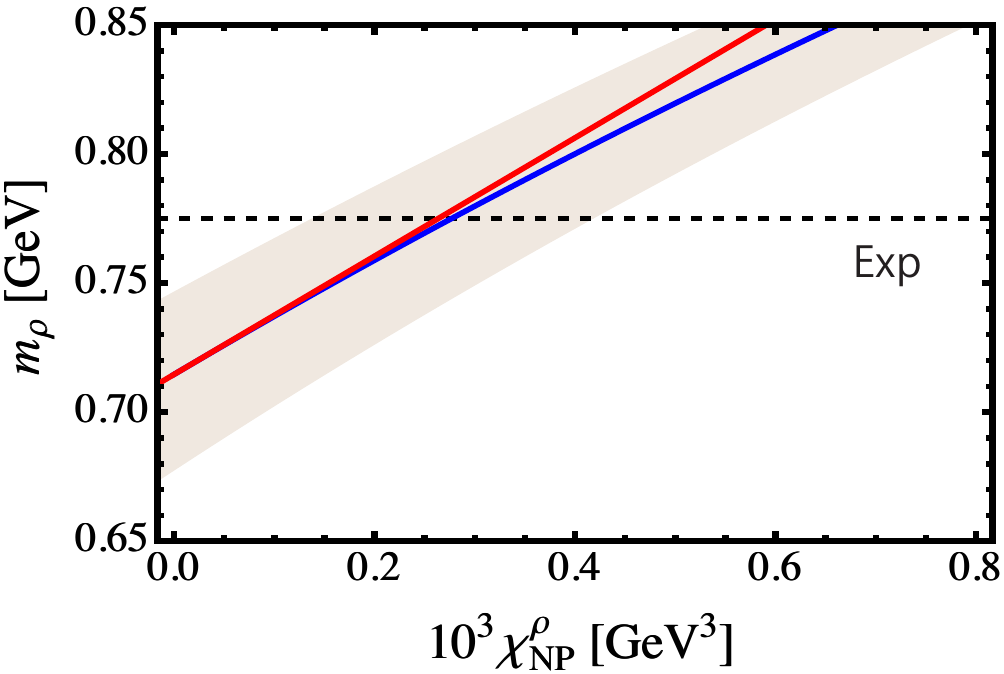}%
\caption{\label{fig1}
 $\rho$-meson mass as a function of $\chi^{\rho}_{\rm NP}$.
 The blue (red) line denotes the numerical solution (linear approximation) for
 $\langle0|\alpha_sG^{a\mu\nu}G^a_{\mu\nu}|0\rangle_{\mathrm{NP}}=7.0\times10^{-2}\;{\rm GeV}^4$. 
 To illustrate the magnitude of the error band,
 we use $\langle0|\alpha_sG^{a\mu\nu}G^a_{\mu\nu}|0\rangle_{\mathrm{NP}}=5.7\times10^{-2}\;{\rm GeV}^4$
 and $\langle0|\alpha_sG^{a\mu\nu}G^a_{\mu\nu}|0\rangle_{\mathrm{NP}}=8.3\times10^{-2}\;{\rm GeV}^4$~\cite{Narison:2011xe}.
 The upper value corresponds to the upper edge of the error band.
}
\end{figure}

 Next, we consider the correction to the mass formula from $\chi^{\rho}_{\rm NP}$.
 In Fig.~\ref{fig1},
 we show the $\chi^{\rho}_{\rm NP}$-dependence of the $\rho$-meson mass with the error band
 with respect to the uncertainty of the gluon condensate~\cite{Narison:2011xe}.
 At the experimental value of the $\rho$-meson mass, $m_{\rho}=775$ MeV,
 the corrections beyond linear order of $\chi^{\rho}_{\rm NP}$ are so small
 that the numerical solution is well reproduced by 
\begin{equation}
  \widetilde{m}_{\rho}\sim m_{\rho}
  +\frac{40\pi^{\frac{3}{2}}\sqrt{2}}{3\sqrt{3}}\frac{\chi^{\rho}_{\rm NP}}
  {\Bigl(\langle0|\alpha_sG^{a\mu\nu}G^a_{\mu\nu}|0\rangle_{\mathrm{NP}}\Bigr)^{\frac{1}{2}}}, 
 \label{mass2}
\end{equation}
 where $m_{\rho}$ is given by Eq.~(\ref{s3}).
 To reproduce the experimental $\rho$-meson mass, we have
 $\chi^{\rho}_{\rm NP}=2.6\times10^{-4}\;{\rm GeV}^3$
 for $\langle0|\alpha_sG^{a\mu\nu}G^a_{\mu\nu}|0\rangle_{\mathrm{NP}}=7.0\times10^{-2}\;{\rm GeV}^4$,
 which is in good agreement with the result of the numerical solution,
 $\chi^{\rho}_{\rm NP}=2.8\times10^{-4}\;{\rm GeV}^3$, 
 as shown in Fig.~\ref{fig1}.

 The mass formula of the $\omega$ meson is obtained from the vector current
 $j^{\mu}_{\omega}=(\bar{u}\gamma^{\mu}u+d\gamma^{\mu}d)/2$.
 As in the case of the $\rho$-meson, at the experimental value of the $\omega$-meson mass,
 $m_{\omega}=783$ MeV~\cite{Beringer:1900zz},
 the corrections beyond linear order of $\chi^{\omega}_{\rm NP}$ can be neglected.
 Then, we have 
\begin{equation}
  \widetilde{m}_{\omega}\sim \widetilde{m}_{\rho}
  +\frac{40\pi^{\frac{3}{2}}\sqrt{2}}{3\sqrt{3}}\frac{\delta_{-}\chi_{\rm NP}}
  {\Bigl(\langle0|\alpha_sG^{a\mu\nu}G^a_{\mu\nu}|0\rangle_{\mathrm{NP}}\Bigr)^{\frac{1}{2}}},
 \label{mass3}
\end{equation}
 with $\delta_{-}\chi_{\rm NP}=\chi^{\omega}_{\rm NP}-\chi^{\rho}_{\rm NP}$. 
 The small mass difference between the $\rho$-meson and the $\omega$-meson
 is given by the flavor mixing term in the generalized susceptibilities
 $\delta_{-}\chi_{\rm NP}=-\int{\rm d}^3x\;\langle0|\bar{u}\gamma^{\mu}u(x)\bar{d}\gamma_{\mu}d(0)
  +\bar{d}\gamma^{\mu}d(x)\bar{u}\gamma_{\mu}u(0)|0\rangle_{\mathrm{NP}}/2$.
 To reproduce the experimental mass difference between the $\rho$-meson and the $\omega$-meson, we have
 $\delta_{-}\chi_{\rm NP}=3.5\times10^{-5}\;{\rm GeV}^3$ 
 for $\langle0|\alpha_sG^{a\mu\nu}G^a_{\mu\nu}|0\rangle_{\mathrm{NP}}=7.0\times10^{-2}\;{\rm GeV}^4$,
 which is smaller by an order of magnitude than the flavor diagonal part
 $\delta_{+}\chi_{\rm NP}=\chi^{\omega}_{\rm NP}+\chi^{\rho}_{\rm NP}$.

 Now, let us consider to utilize the sum rule with $s^2$ weight
 instead of our sum rule with $\sqrt{s}$ weight~\cite{Krasnikov:1982ea}. 
 By taking only the four-quark condensates into account, 
 we have $ s_{\rho0,\omega0}= 2m_{\rho,\omega}^2$, $F_{\rho,\omega}= m_{\rho,\omega}^2/(2\pi)$, and
 $m_{\rho,\omega}^6=48\pi^3\alpha_s\langle0|(\bar{u}\gamma^{\mu}\gamma^5t^au\mp \bar{d}\gamma^{\mu}\gamma^5t^ad)^2/2
  +(\bar{u}\gamma^{\mu}t^au+ \bar{d}\gamma^{\mu}t^ad)\sum_{q=u,d,s}\;\bar{q}\gamma_{\mu}t^aq/9
  |0\rangle_{\mathrm{NP}}$.
 Here, $t^a$ are defined as $\lambda^a/2$ with $\lambda^a$ being the Gell-Mann matrices. 
 The coupling constant $g_{\rho}$ is given by $g_{\rho}^2/(4\pi)=\pi$,
 which is consistent with our result $g_{\rho}^2/(4\pi)=8\pi/9$.
 To evaluate the four-quark matrix elements, we retain only the vacuum intermediate state~\cite{Shifman:1978bx}.
 Then, the four-quark operators are reduced to the square of the chiral condensate
 $\langle0|\bar{q}q|0\rangle\equiv\langle0|\bar{u}u|0\rangle=\langle0|\bar{d}d|0\rangle$,
 and we have
 $m_{\rho,\omega}\sim2.8|\sqrt{\alpha_s}\langle0|\bar{q}q|0\rangle_{\mathrm{NP}}|^{1/3}$~\cite{Krasnikov:1982ea}. 
 Here, the flavor mixing condensates such as
 $\langle0|\bar{u}\gamma^{\mu}t^au\bar{d}\gamma_{\mu}t^ad|0\rangle_{\rm NP}$ are neglected, so that 
 the masses of the $\rho$ meson and $\omega$ meson are degenerate.
 In our approach, 
 it corresponds to neglecting $\delta_{-}\chi_{\rm NP}$. 
 The vacuum saturation Ansatz, which is exact in the large-$N_{\rm c}$ limit,
 might underestimate the four-quark condensates and
 causes the ambiguities of the absolute values of the masses of the vector mesons.
 On the other hand, in our approach,
 we can estimate them from the gluon condensate without factorization Ansatz.
 For the gluon condensate, the consistent values are estimated
 from several approaches~\cite{Narison:2002pw},
 so that it enables us to estimate the masses of the vector mesons with small error bands.

 Finally, let us consider the axial-vector current with quantum numbers of the $a_1$-meson:
 $j^{\mu}_{a_1}=(\bar{u}\gamma^{\mu}\gamma^5u-d\gamma^{\mu}\gamma^5d)/2$.
 To obtain simple mass formula of the $a_1$-meson,
 we follow the Weinberg's sum rules approach and adopt the following Ansatz~\cite{Weinberg:2005kr}:
\begin{equation}
 \rho_A(s)=\pi f_{\pi}^2\delta(s)+F_{a_1}\delta(s-m_{a_1}^2)+\rho^{\rm pert.}_V(s)\Theta(s-s_{\rho0}) ,
 \label{a1Ansatz}
\end{equation} 
 with $f_{\pi}$, $m_{a_1}$ and $F_{a_1}$ being the pion decay constant $f_{\pi}=130$ MeV,
 the mass and the pole residue, respectively. 
 Here, by assuming the chiral invariance of the continuum,
 the continuum part of the spectral function of the axial-vector current
 is taken to be equal to the vector current's one.
 By using our result $F_{\rho}=9m_{\rho}^2/(16\pi)$ 
 instead of the vector meson dominance used in the original Weinberg's work,
 we obtain the $a_1$-meson mass from
 the first and second Weinberg's sum rules with $s^{0,1}$ weights:
\begin{eqnarray}
  F_{a_1}&=&  \frac{9}{16\pi}m_{\rho}^2\Bigl(1-\Bigl(\frac{4\pi f_{\pi}}{3m_{\rho}}\Bigr)^2\Bigr) ,
 \label{a1} \\
  m_{a_1}^2&=&\frac{m_{\rho}^2}{1-\Bigl(\frac{4\pi f_{\pi}}{3m_{\rho}}\Bigr)^2},
 \label{a2}
\end{eqnarray}
 where $m_{\rho}$ is given by Eq.~(\ref{s3}).
 We have $m_{a_1}=1.1$ GeV
 for $\langle0|\alpha_sG^{a\mu\nu}G^a_{\mu\nu}|0\rangle_{\mathrm{NP}}=7.0\times10^{-2}\;{\rm GeV}^4$,
 which is smaller only by $10$\%
 than the the experimental value, $m_{a_1}=1.23$ GeV~\cite{Beringer:1900zz}. 
 Also, we have $\chi^{a_1}_{\rm NP}$ from our sum rule with $\sqrt{s}$ weight:
\begin{equation}
 \chi^{a_1}_{\rm NP}=-\frac{27}{64\pi^2}\Biggl(1-\sqrt{1-\Bigl(\frac{4\pi f_{\pi}}{3m_{\rho}}\Bigr)^2}\Biggr)m_{\rho}^3.
 \label{a3}
\end{equation}
 If we take $\langle0|\alpha_sG^{a\mu\nu}G^a_{\mu\nu}|0\rangle_{\mathrm{NP}}=7.0\times10^{-2}\;{\rm GeV}^4$,
 we obtain $\chi^{a_1}_{\rm NP}=-5.5\times10^{-3}\;{\rm GeV}^3$.
 $\chi^{a_1}_{\rm NP}$ becomes negative as clearly seen in Eq.~(\ref{a3}),
 and its absolute value is larger by an order of magnitude than
 the one of $\chi^{\rho}_{\rm NP}$.

 It is straightforward to apply our approach to the currents with valence strange quarks.
 Since the strange quark mass is comparable to $\Lambda_{\rm QCD}$,
 we may need to treat the strange quark mass carefully.
 However, this is left for future work.

 \section{Concluding remarks}

 In this paper, we have derived the mass formulas
 for the vector and axial-vector mesons written in terms of
 the gluon condensate on the basis of finite energy QCD sum rules. 
 Our starting point is the spectral decomposition of the current correlation function.
 By considering energy-weighted integrations of the spectral function,
 we obtained equations
 which related the spectral sums with $s^{n}$ and $s^{n+1/2}$ weights  
 to the vacuum expectation values of commutation relations
 between the current and the Hamiltonian of QCD.  
 Both sides of the sum rules were UV divergent, so that 
 the subtraction of them was necessary to obtain finite sum rules. 
 We found that the sum rules with $s^{n}$ weights were reduced to
 the ones obtained from OPE. 
 On the other hand, the sum rules with $s^{n+1/2}$ weights could not be obtained from OPE.
 They gave new relations between the spectral sums and QCD vacuum fluctuations.

 By adopting the standard pole $+$ continuum Ansatz
 for the spectral functions of the vector currents, 
 we solved the sum rules with $s^{0,1/2,1}$ weights,
 and estimated the masses of the vector mesons 
 from the gluon condensate without the vacuum saturation Ansatz.
 We utilized our new sum rule with $s^{1/2}$ weight
 instead of the sum rule with $s^2$ weight
 usually used in the standard QCD sum rules approach.
 The corrections to the masses from the generalized susceptibilities would be small. 
 Therefore, the estimates with retaining only the gluon condensate
 were in rather good agreement with the experimental masses.
 We have discussed the $a_1$-meson mass
 from the first and second Weinberg's sum rules with $s^{0,1}$ weights.
 We assumed the chiral invariance of the continuum parts of the spectral functions,
 and obtained the $a_1$-meson mass formula
 written in terms of the gluon condensate and the pion decay constant,
 which was in good agreement with the experimental $a_1$-meson mass. 
 Also, we estimated the generalized susceptibilities
 in the vector and axial-vector currents. 
 We found that they have opposite signs and 
 the absolute value in the axial-vector current
 is larger by an order of magnitude
 than the one in the vector current.
 Investigations of them from lattice QCD simulations
 are necessary to confirm the reliability of our analysis presented in this paper.

 We have several future directions to extend our analysis presented in this paper.
 Mass formulas for other mesons and baryons,
 and applications of our approach to in-medium QCD sum rules~\cite{Hayano:2008vn} 
 would be the important future problems. 
 It is also interesting future work
 to discuss the phenomenological application of the new sum rules with $s^{n+1/2}$ weights.
 For example, those sum rules may also be used to reconstruct the spectral function
 on the basis of the Bayesian technique such as the Maximum Entropy Method~\cite{Gubler:2010cf}.

\begin{acknowledgements}

 The author thanks T.~Hatsuda, Y.~Hidaka, S.~Sasaki and A.~Yamamoto
 for stimulating discussions and helpful comments.
 T.~H. is supported by JSPS Research Fellowships for Young Scientists.

\end{acknowledgements}

\end{document}